\documentclass[%
aip,
 amsmath,amssymb,
reprint,%
]{revtex4-1}
\pdfoutput=1
\usepackage{graphicx}
\usepackage{dcolumn}
\usepackage{bm}
\usepackage{float}
\usepackage{amssymb}
\begin{document}

\preprint{}

\title[]{Heterogeneous nucleation of pits via step pinning during Si(100) homoepitaxy}


\author{E. N. Yitamben}
\affiliation{Sandia National Laboratories, PO Box 5800 Albuquerque New Mexico, 87185  USA}

\author{R.E. Butera}
\affiliation{Laboratory for Physical Sciences, 8050 Greenmead Drive, College Park, MD 20740}

\author{B. S. Swartzentruber}
\affiliation{Center for Integrated Nanotechnologies, Sandia National Laboratories, Albuquerque New Mexico, 87185  USA}

\author{R. J. Simonson}
\affiliation{Sandia National Laboratories, PO Box 5800 Albuquerque New Mexico, 87185  USA}

\author{S. Misra}
\affiliation{Sandia National Laboratories, PO Box 5800 Albuquerque New Mexico, 87185  USA}

\author{M. S. Carroll}
\affiliation{Sandia National Laboratories, PO Box 5800 Albuquerque New Mexico, 87185  USA}

\author{E. Bussmann}
\affiliation{Center for Integrated Nanotechnologies, Sandia National Laboratories, Albuquerque New Mexico, 87185  USA}

\date{\today}

\begin{abstract}

Using scanning tunneling microscopy (STM), we investigate oxide-induced growth pits in Si thin films deposited by molecular beam epitaxy. In the transition temperature range from 2D adatom islanding to step-flow growth,  systematic controlled air leaks into the growth chamber induce pits in the growth surface.  We show that pits are also correlated with oxygen-contaminated flux from Si sublimation sources.  From a thermodynamic standpoint, multilayer growth pits are unexpected in relaxed homoepitaxial growth, whereas oxidation is a known cause for step-pinning, roughening, and faceting on elemental surfaces, both with and without growth flux. Not surprisingly,  pits are thermodynamically metastable and heal by annealing to recover a smooth periodic step arrangement. STM reveals new details about the pits' atomistic origins and growth dynamics. We give a model for heterogeneous nucleation of pits by preferential adsorption of \AA-sized oxide nuclei at intrinsic growth antiphase boundaries, and subsequent step pinning and bunching around the nuclei. 

\end{abstract}
\pacs{68.35.B-, 68.35.bg, 68.37.Ef, 68.55.A-, 68.55.ag, 68.55.J-, 68.35.Md}
\keywords{thin film, molecular beam epitaxy, scanning tunneling microscopy, Si(100), oxide, growth pit, defect, impurity}
\maketitle

\section{\label{sec:level1} Introduction} 
Evolving Si quantum computing technologies are demanding epitaxial thin films with exceptionally flat and abrupt interfaces, ideally to the single-atomic-layer limit.~\cite{z12, R16, G07,m9, Fr10} For a few leading qubit platforms, {\it e.g.} e$^-$ spin qubits in quantum dots in strained-Si/SiGe and atomic-precision Si:P donors, the need for smooth abrupt interfaces stems from various difficulties originating from interplay of roughness and the multivalley physics of the Si conduction band.~\cite{k01,Fr10,C10,JK15}  

Apart from anticipated kinetic roughness via 2D island nucleation at lower growth temperatures, anisotropic diffusion and step-attachment specific to the Si(100)-$2\times1$ surface introduce diverse roughening instabilities in Si vapor-phase growth via molecular beam epitaxy (MBE) and chemical vapor deposition (CVD).~\cite{Ha89,Mo89,Ea90,E97,k97,P2000,Z1,v01} Numerous previous works have successfully linked mesoscopic roughness to growth kinetic instabilities rooted in anisotropies of the $2\times1$ surface.~\cite{WU93,v01,S98, M22,M10} A common feature of such instabilities is that they involve kinetic step-bunching to form ripples, striations, and bumps in the growth surface, often evolving characteristic  \{113\} facets.~\cite{H93,O95}  A postulated Ehrlich-Schwoebel (E-S) barrier at double-height rebonded B-type steps (denoted D$_{B}$)~\cite{EBB17} where the dimer bond is parallel to step edge, plays a role in many models for such instabilities and \{113\} facets eventually evolve from D$_{B}$ step bunches.~\cite{H93,O95} 

Works going back to the earliest studies of vapor-phase crystal growth describe multilayer step pinning growth pits in Si under various deposition conditions, both in MBE and CVD.~\cite{Jo67,Jo68,Ab68, C71,Me86,Ka96,Wi97, D98,J2,Ga9,M10s,Ar13} Intuitively, spontaneous pit formation in homoepitaxy is unlikely without some driving force, such as bulk strain relief.~\cite{Te94, V98} It is likely that pits are a metastable feature caused by step pinning and bunching around extrinsic defects, {\it e.g.}  particles adhered to the surface~\cite{Me86}, and some previous studies correlated step-pinning and pitting effects with extrinsic nuclei formed of oxide or carbon.~\cite{Ab68, C71, Ka96, D98}

By contrast, a few recent works examining the atomic-scale details of growth pit formation attempt to explain pits as a spontaneous intrinsic process in the islanding growth mode, T$\lesssim 600^{\circ}$C.~\cite{M10s, Ar13} For example, one model~\cite{M10s} attributes pit formation to step-pinning and bunching at a kink defect on antiphase boundaries (APBs) intrinsic to Si epitaxy in the islanding growth mode.~\cite{Ha89,Br93} The pits deepen by step bunching stabilized by the postulated E-S barrier at D$_{B}$ steps, and then eventually evolve \{113\} facetted sidewalls.  
 
Here, we report STM studies of Si thin film growth revealing details of growth-pit formation during Si homoepitaxy that point to a different conclusion. Specifically, contrary to recent studies,~\cite{M10s, Ar13} we find that oxygen plays an important role in growth pit nucleation. We find that pits are not a consistent reproducible feature of epitaxy because they are induced by oxide contamination caused by air leaks or oxygen-rich flux from contaminated Si cells. Pits initiate at \AA - sized step-pinning oxide defects that adsorb preferentially at APBs between 2D Si adatom islands. The oxide defects nucleate pits by preventing completion of the subsequent atomic layers, plausibly because Si-Si interlayer recoordination from surface $2\times1$-dimerization to tetrahedral bulk bonding is blocked by stronger Si-O bonds. Pits deepen by subsequent step bunching around pit nuclei.  A steady-state growth condition is reached when the surface is so dense with pits that the remaining terraces support only single nucleation events, thereby eliminating new APBs, thereby slowing oxide adsorption and pit nucleation and enforcing a nucleation-limited smoothing.  Oxygen-induced pits are thermodynamically metastable and fill-in during short anneals hot enough to desorb the oxide pinning centers.  

\section{\label{sec:level1} Methods and materials}

We use Si(100) substrates ($2\times10^{16}$ cm$^{-3}$ boron) with  $< 0.1^{\circ}$ miscut tilted toward the (010) direction. The miscut results in terraces $100-150$ nm wide separated by single-height atomic steps. Owing to the (010) tilt, each step is composed of equal portions of A and B- type edge, and adjacent steps are indistinguishable. 

Prior to experiments, the substrates are prepared by a wet chemical cleaning procedure that includes three cycles of chemical oxidation ($3:1$ H$_2$SO$_4:$H$_2$O$_2$, 90$^{\circ}$C, 5 mins) and reduction ($6:1$ NH$_4$F:HF, 10s), followed by a rinse in deionized H$_2$O. The final step is an oxidation ($5:1:1$ H$_2$O:H$_2$O$_2$:HCl, 60$^{\circ}$C, 5 mins). The substrates are then loaded into the ultrahigh vacuum (UHV) STM systems.

 In UHV, clean $2\times1$-reconstructed surfaces are prepared by direct current self-heating the substrate at $1200^{\circ}$C for 30s to drive off residual gas and oxide, then cooling to 600$^{\circ}$C for several minutes, and then annealing again at $1200^{\circ}$C for 10 s. Fig.~\ref{fig:fig1} (a) shows a typical example of the clean step-terrace structure after annealing. The surface is $2\times1$-reconstructed, as shown in subsequent figures, with defect densities $< 0.01$ (intrinsic dimer vacancy and C-type defects).

We performed growth studies in two separate UHV STM systems equipped with solid Si sublimation sources. One vacuum system is homebuilt and equipped with a homemade Si sublimation source comprised of a Joule-heated Si chip fitted between Mo/Ta mounts, while the other system is commercially available with a commercial sublimation source.~\cite{Eb17}   

We use growth rates ranging from 0.1-0.5 \AA/s. This range of growth rates is impractically slow for many Si device-layer growth applications, but certain quantum device research applications demand similar low growth rates.~\cite{R16} Both systems grow phosphorus-doped Si (N$_D\sim5\times10^{17}$cm$^{-3}$). The deposition sources are degassed and conditioned by running hot for at least 12 hours at sublimation (T$\gtrsim 1100^{\circ}$C) amounting to more than a micrometer of sublimated Si. In the step-flow growth regime for T$\gtrsim650^{\circ}$C, both systems consistently produce smooth flat epitaxy with typical contaminant (O, C, and N) concentrations $< 10^{18}$ cm$^{-3}$.  As the growth temperatures are reduced, the two growth systems produce qualitatively varying surface morphologies to be described in this paper.

During growth, T is measured by pyrometer (Process Sensors Metis MP-25) calibrated by H desorption (T$=480^{\circ}$C) from the monohydride H:Si(100)$-2\times1$ phase. In growth studies, we estimate that the uncertainty in T is $\pm 50^{\circ}$C at T$=250^{\circ}$C dropping to $\pm 10^{\circ}$C above T$\sim600^{\circ}$C. Owing to variability and spatial nonuniformity of the self-heating process, there is a similar $\pm 50^{\circ}$C variation spatially across the samples. 

When sitting idle, the growth chamber pressures are 7$\times$10$^{-11}$ and 5$\times$10$^{-10}$ torr, in the commercial and homebuilt systems, respectively. In both systems the background pressures rise to 1-2 $\times$10$^{-9}$ torr during growth cycles, primarily owing to increased ambient H$_2$. The gas composition in the growth chamber is measured using a residual gas analyzer (RGA) (SRS RGA100) mounted about 20 cm from the sample. 

\section{\label{sec:level1} Results}

\subsection{\label{sec:level2} Clean Si thin-film growth characteristics}

\begin{figure*}
 \includegraphics[width=7 in]{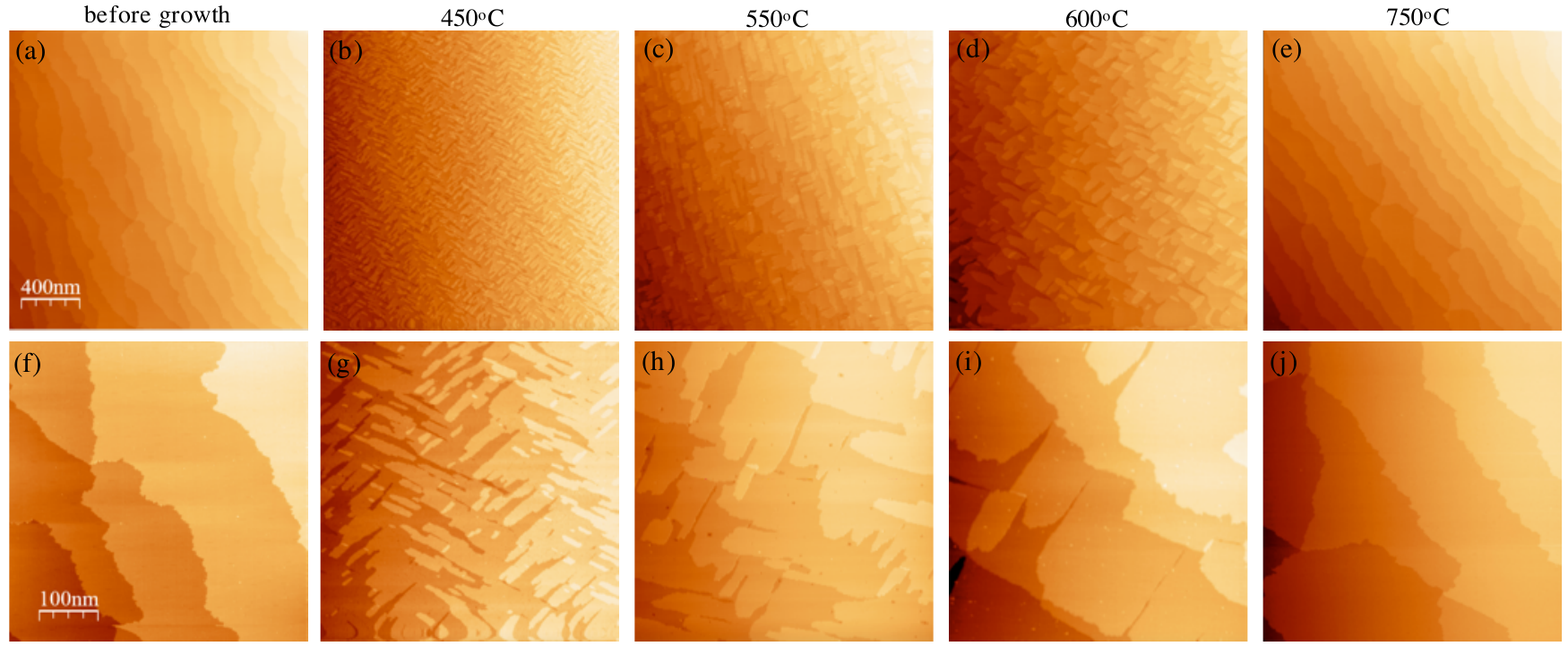}%
\caption{Characteristic Si growth surfaces versus T. The top row shows $2\times2\mu$m$^2$ images of the overall step terrace characteristics, and the bottom row shows  $0.5\times0.5$ $\mu$m$^2$ images revealing shape and density of islands during growth. The vertical scale is evident from single atomic layer steps, height $=a/4 =1.36$~\AA, visible throughout each image.  Although the film thickness (50, 8, 100, and 3 nm from left to right) varies between images, the step-terrace structure and island density are thickness-invariant, {\it i.e.} growth is near steady-state, for the chosen growth rates in the range 0.2-0.4 \AA/s. These filled-states STM images were acquired with biases in the range -1.9 V to -3.2 V at $I=100$ pA.}%
\label{fig:fig1} 
\end{figure*}

As a baseline for studies in this paper, we start by describing nanoscale traits of ideal epitaxial Si growth surfaces over the temperature transition (450$\lesssim $T$\lesssim 750^{\circ}$C) from islanding to step-flow growth produced in our homebuilt system. For Si device applications, growth temperatures are typically chosen in this range.~\cite{R16} Fig.~\ref{fig:fig1} shows STM images of growth surfaces as a function of T. We see only generally anticipated epitaxial features that are common to all of our thin films: (1) for T$\lesssim 650^{\circ}$C, 2D adatom island nucleation is routine, Fig.~\ref{fig:fig1} (b-d), while (2) for  T$\gtrsim 650^{\circ}$C, we repeatably observe only step-flow growth, Fig.~\ref{fig:fig1} (e), characterized by periodic trains of single monolayer atomic steps.  Our results are qualitatively consistent with results described in many previous studies and growth models.~\cite{Z97,v01} The temperature for step-flow growth depends weakly on miscut (terrace width). Generally, adatom islands become sparse with increasing temperature and vanish as surface diffusion lengths become large with respect to the distance between adjacent atomic steps.~\cite{Mo89,v01}

\subsection{\label{sec:level2} Growth pits induced by an air leak}

\begin{figure*}
 \includegraphics[width=7in, angle=0]{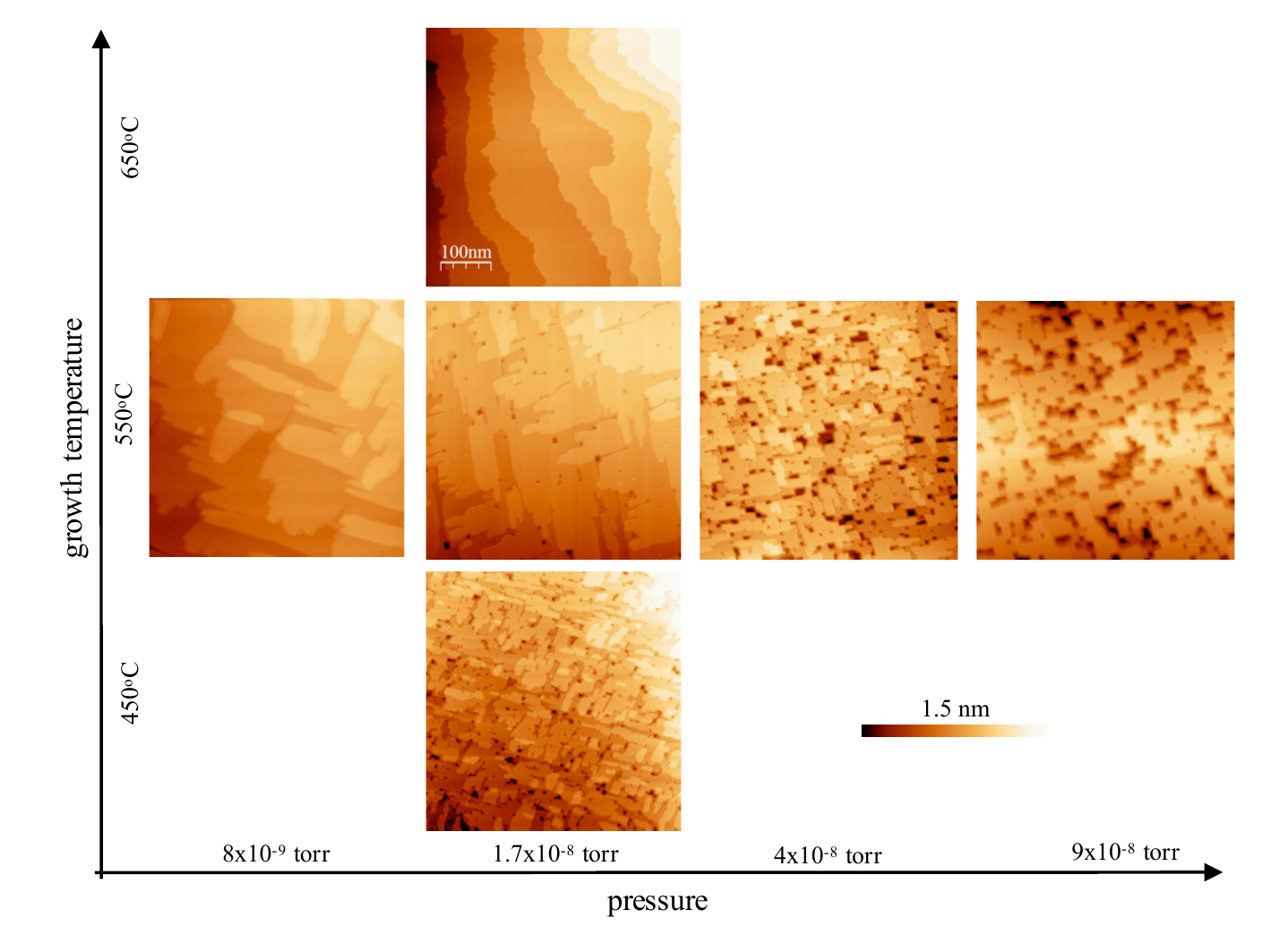}%
\caption{Air-induced pit formation versus substrate temperature and pressure (air $+$ UHV background) during Si deposition.  These 8 nm-thick films were grown at $0.2$ \AA/s. The images are $500\times500$ nm$^2$, and they were acquired in filled states at biases of $-2$ to $-3$V at  $I=100$ pA. }%
\label{fig:fig2} 
\end{figure*}

A controlled air leak into the growth chamber during Si deposition induces a new feature, nanosized pits, in the growth surface,  Fig.~\ref{fig:fig2}. The pitted morphology was obtained in the homebuilt system in growth cycles interleaved with the clean epitaxy in Fig.~\ref{fig:fig1}.  

The pits are (110)-oriented four-fold symmetric openings comprised of dense step bunches. The corners of the pits tend to round with increased T and growth thickness, since both T and pit size increase the probability for kinks in steps bounding the pits.  The smallest and shallowest pits are discernible from regular equilibrium surface vacancy defects when they exceed 2-3 ML depth and roughly 1-2 nm on a side. The deepest pits approach but do not exceed the film thickness, suggesting that the pits are caused by step-pinning at some defect site introduced during film deposition. 

Fig.~\ref{fig:fig2} shows the increasing pit density versus air pressure $>10^{-8}$ torr. The pit density diminishes and vanishes for T$\gtrsim 650^{\circ}$C, as step-flow growth becomes dominant. The T-dependence of pit density suggests that there is a correlation between 2D island nucleation and pit formation. 

\begin{figure}
 \includegraphics[width=250pt, angle=0]{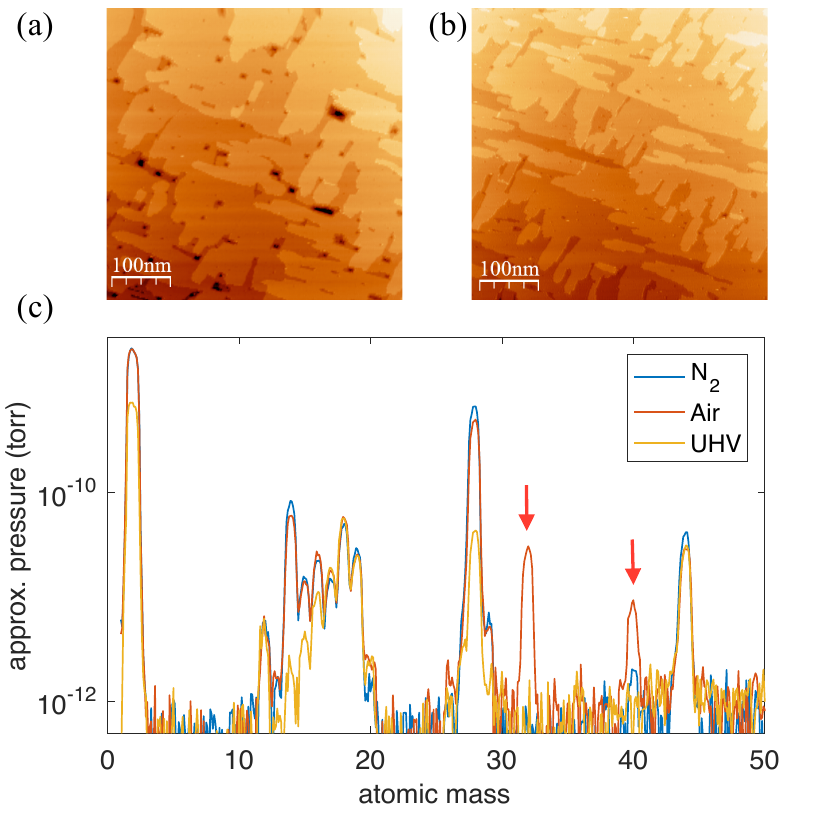}%
\caption{Growth surfaces of 8 nm-thick films deposited with background pressures of $1.7\times10^{-8}$ torr ambient (a) air {\it versus} (b) N$_2$. The growth conditions are T$=550^{\circ}$C and deposition rate $0.2$ \AA/s. These images were acquired in filled-states at -2.8 V at I$=100$ pA. (c) RGA of the ambient during the growths in the upper panel. Arrows indicate molecular oxygen and argon in air. }%
\label{fig:fig3} 
\end{figure}

By contrast, leaking high purity N$_2$ into the growth chamber at equivalent pressures does not induce nearly as large a density of pits.  Fig.~\ref{fig:fig3} compares depositions performed with air {\it versus} high purity N$_2$ leaks. It is likely that some other constituent of air, besides N$_2$, is responsible for pit formation. By comparing the RGA data in Fig.~\ref{fig:fig3} (c), the only detectable significant differences are partial pressures of O$_2$, and Ar (mass 32 and 40). At the growth temperatures here, molecular oxygen has a sticking coefficient, S$_o = 0.02 -0.05$, and well-characterized oxidation reactions with the surface.~\cite{Av91,Se94,Se95}
 
\subsection{\label{sec:level2} Growth pits caused by contaminated growth flux}

We find that oxygen-contaminated flux from Si sublimation sources also induces growth pit formation. Residual oxides in or on a sublimation source provide a detectable flux of oxygen-bearing species, causing both pits and bulk oxygen contamination.

\begin{figure}
 \includegraphics[width=250pt, angle=0]{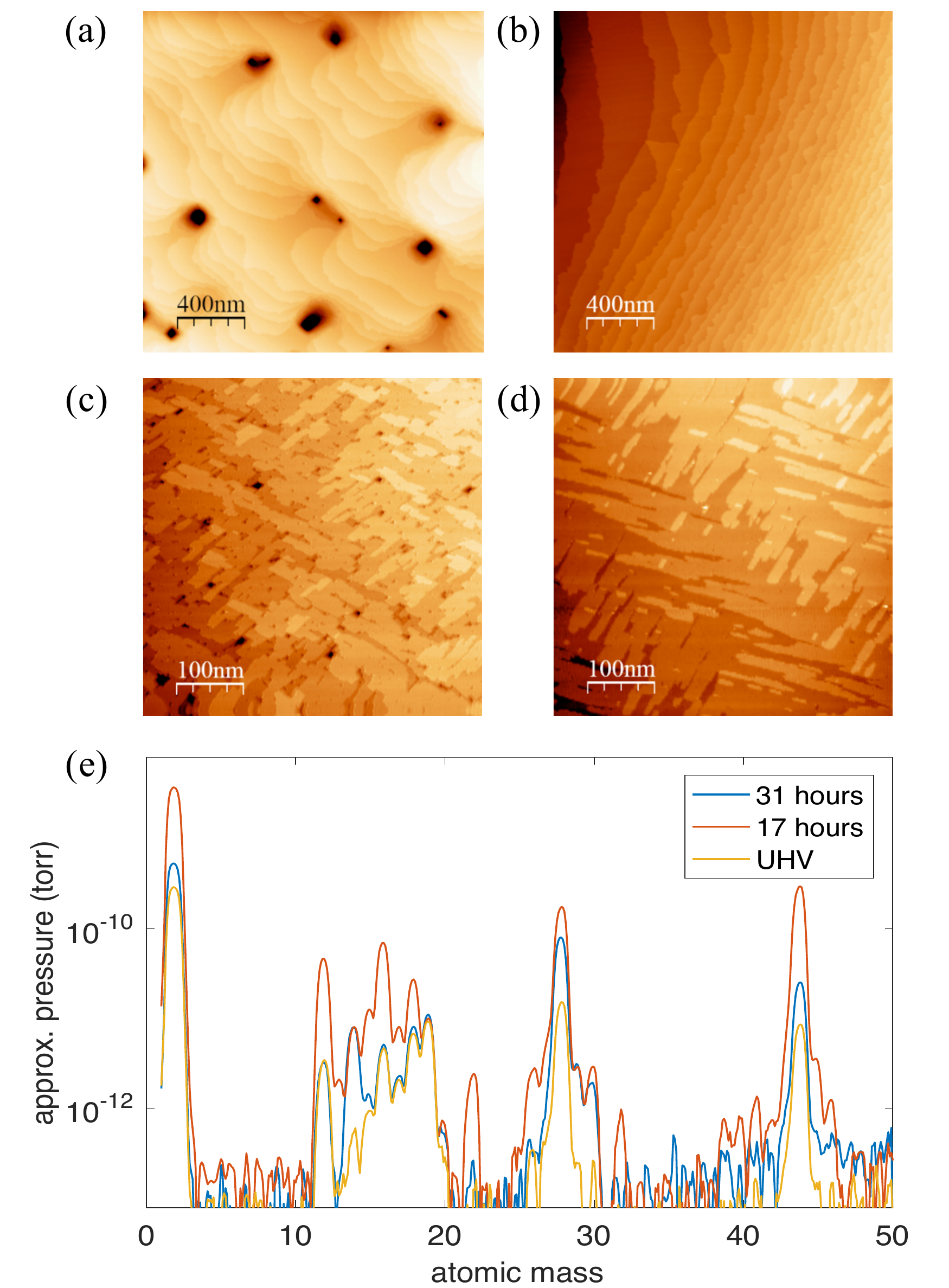}%
\caption{ Typical growth surfaces produced by (a) the commercial sublimation source on a 20 nm-thick film grown at T$=600^{\circ}$C and $0.2$ \AA/s and (b) the commercial sublimation source for a 10 nm-thick film grown at T$=800^{\circ}$C and $0.2$ \AA/s. Typical growth surfaces produced by (c) the homebuilt source after inadequate degassing, and (d) longer degassing of the source material. These are 8 nm-thick films grown at T$=450^{\circ}$C at a rate of $0.2$ \AA/s. The images, which are 2$\times2\mu$m$^2$ (upper panels) and 500$\times$500 nm$^2$and were acquired in filled-states at -2 to -3 V at I$=100$ pA. (e) RGA spectra during source degassing showing how oxygen-related contamination peaks diminish as the deposition source and material are degassed.}%
\label{fig:fig4} 
\end{figure}

Thru a few years of experience, it has become clear that our commercial Si cell frequently (not always) produces pitted thin films.
Fig.~\ref{fig:fig4} (a) and (b) show examples of typical growth surfaces produced by the commercial source. 
The pitted morphology occurs only up to around T$\sim 650^{\circ}$C, giving way to smooth step-flow growth at higher temperatures, Fig.~\ref{fig:fig4} (b).
The variation between pitted {\it vs} smooth growth morphology suggests that pitting is driven by a secondary uncontrolled variable, which we will demonstrate is most likely residual oxygen contamination.

If degassed inadequately, our homebuilt source produces thin films with pitted surfaces too, but the pit density diminishes as a function of cumulative deposition time as the Si source material is  conditioned.  Fig.~\ref{fig:fig4} (c) and (d) show growth surfaces produced by the homebuilt source versus source burn-in time. Fig.~\ref{fig:fig4} (c) shows a typical pitted growth surface produced after roughly 17 hours of source conditioning via sublimation ($\sim$1100$^{\circ}$C, sublimation of a few micrometers of Si), while (d) shows a smooth pit-free growth with fourteen more hours of conditioning.

The correlation between the pitting in Fig.~\ref{fig:fig4} (c) and oxygen-bearing contaminants is elucidated by RGA spectra sampling the growth flux. 
Typical RGA spectra are shown in Fig.~\ref{fig:fig4} (e). After 17 hours of degassing, we see Si peaks (mass 28-30) along with a variety of peaks consistent with additional residual oxygen, including small atomic (16) and molecular oxygen peaks (mass 32), elevated water (18) and SiO (masses 44-46). Whereas, after 31 hours of source degassing, there is no longer detectable molecular oxygen, SiO, or additional water above the UHV background. 

Consistent with oxygen-rich flux gas, both of our sources tend to produce oxygen-rich thin films as the substrate growth temperature is reduced. According to secondary ion mass spectroscopy (SIMS) for growth temperatures T$\lesssim 300^{\circ}$C, both the commercial and homebuilt system typically produce epitaxy with residual oxygen concentration on the order of $10^{19}$cm$^{-3}$.

\subsection{\label{sec:level2} Pit nucleation versus T}

 To get insight into the atomistic origin of the pits, we have performed a T-dependent study of the early stages of film growth that reveals a correlation between 2D adatom island nucleation and pit formation. For both air-induced pits, and pits correlated with contaminated source flux, the pit density diminishes rapidly with growth temperature, Fig.~\ref{fig:fig2} and Fig.~\ref{fig:fig4}. Pits vanish at the step-flow transition. In this study, we used the homebuilt source in an inadequately degassed (16h at 1200$^{\circ}$C) condition, that produced epitaxy similar to Fig.~\ref{fig:fig4} (c). The various pit sizes in Fig.~\ref{fig:fig4} (c) suggests that pits nucleate at various times during the evolution of the film, and we found it necessary to grow at least 2 nm-thick Si ($\sim 16$ monolayers) in order to see the first easily identifiable few-monolayer pits in a reasonably-sized STM image (we chose 500$\times$500 nm$^2$).  

\begin{figure}[h]
 \includegraphics[width=240 pt, angle=0]{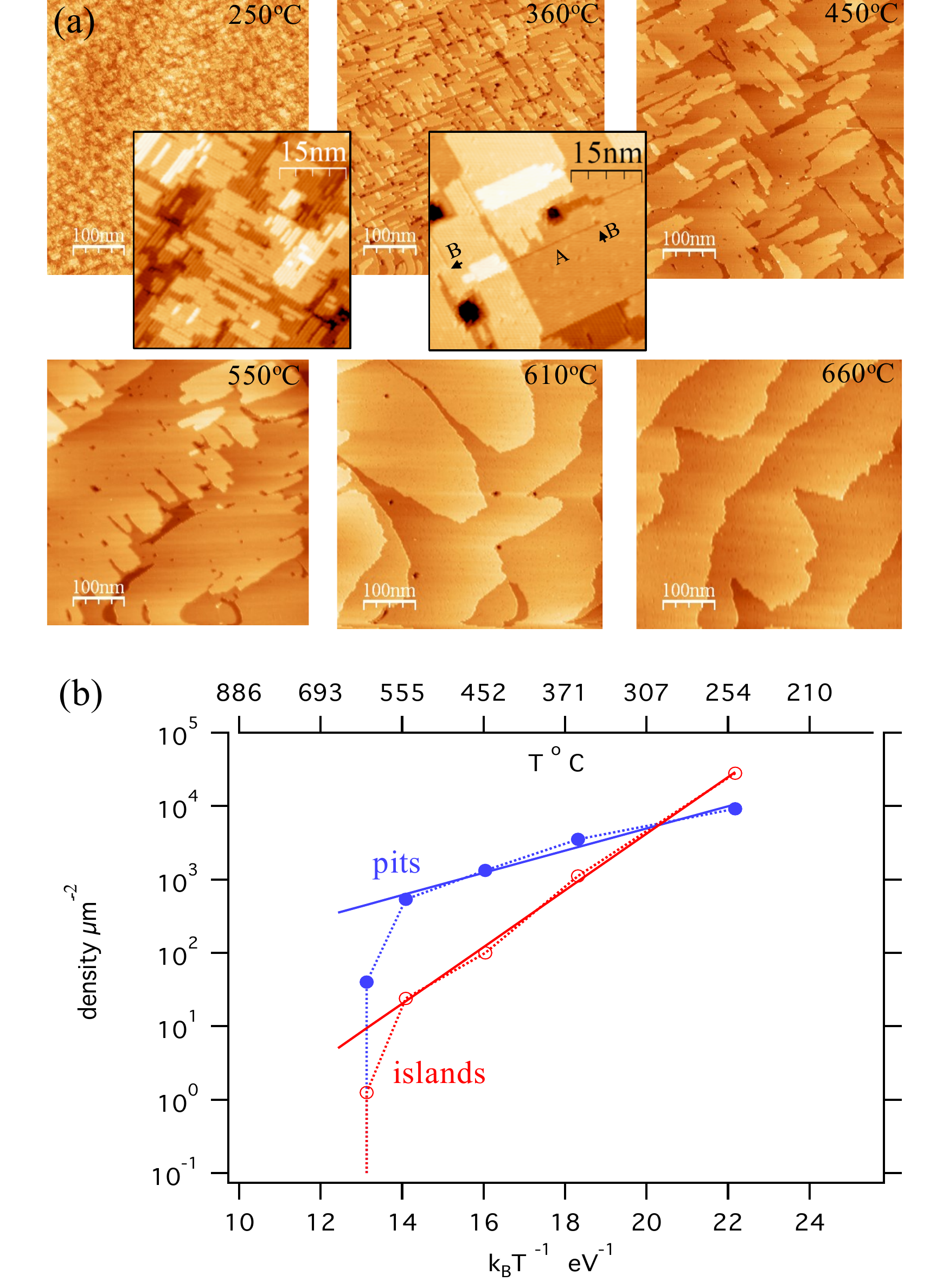}%
\caption{(a) STM images of thin film topography as a function of T for films deposited at $0.2$ \AA/s to a thickness of 3$\pm$0.4 nm. These $500\times500$ nm$^2$ images (insets are $45\times45$ nm~$^2$) were acquired in filled-states at -2 to -3 V at I$=100$ pA. A few growth APBs are indicated by lettering A or B and arrows. (b) Temperature dependence of the pit and island densities versus T. The circles connected by dashed lines are the measured data, and solid lines are fits to the first four data points from the right. The island density is only approximate for $T<450^{\circ}$C owing to island coalescence. }%
\label{fig:fig5} 
\end{figure}

The results of the growth study are shown in Fig.~\ref{fig:fig5}. At growth $T= 250^{\circ}$C, surface roughness is so dominated by small 3D island stacks that it is hard to identify definitive pits. This characteristic island-upon-island stacking mode is a special consequence of growth dynamics on the $2\times1$ surface, where islands in the topmost layer nucleate at the growth APBs in the layer below.~\cite{Ha89, Z1, v01}  For T=$360^{\circ}$C, isolated pits with depth $\> 2$ atomic layers are clear. It is clear that the pit density diminishes rapidly with increasing T, reaching zero for 610$<$T$<660^{\circ}$C.

 We have plotted the pit density {\it vs.} T in Fig.~\ref{fig:fig5} (b), along with the estimated island density. For $T<610^{\circ}$C, pit and island densities diminish roughly exponentially with increasing T. The pit density diminishes as $\exp(0.35\pm0.1 \textrm{eV}/\textrm{k}_B \textrm{T})$, while the islanding density is $\exp(0.9\pm0.1 \textrm{eV}/\textrm{k}_B \textrm{T})$. Both the pit and island densities diverge from the exponential trend and vanish abruptly for  610$<$T$<$660$^{\circ}$C.  Islands disappear as step-flow growth becomes dominant. The simultaneous extinction indicates that pit formation is related to 2D islanding, consistent with findings of Xu {\it et al.}~\cite{M10s}

\subsection{\label{sec:level2} Details of pit growth}

This section describes the evolution of individual pits, from nucleation, then subsequent pit growth by step pinning and bunching, and the consequent evolution of the surface as pits become dense.

\begin{figure}[h!!]
 \includegraphics[width=200 pt]{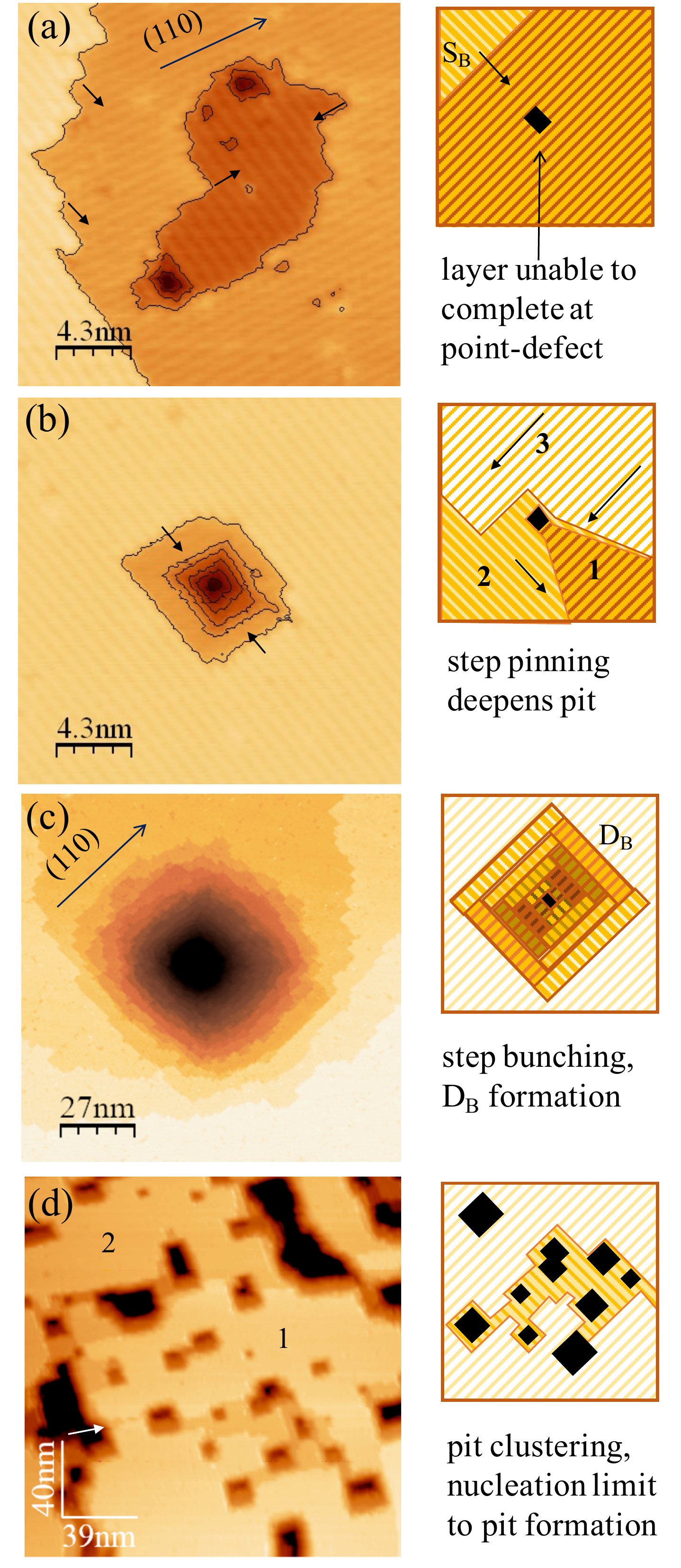}%
\caption{STM images of pit nucleation and stages of growth. (a) A STM image of two pits only 2 layers deep. Arrows on S$_{B}$ steps indicate the direction of advance during growth. (b) A 6 monolayer-deep pit (growth $T=550^{\circ}$C at $0.2$ \AA/s,  contaminated source). (c) Steps bunch and double to form D$_{B}$ steps around an isolated pit several nanometers deep (growth $T=600^{\circ}$C at $0.2$ \AA/s,  contaminated source). (d)  longer time evolution to pit complexes separated by flat ledges. Only two atomic layers, indicated by number, dominate ledges between pits. The arrow points-out a growth boundary (growth $T=550^{\circ}$C at $0.2$ \AA/s,  in $9\times10^{-8}$ torr air). Panels to the right describe evolution of pits and the surrounding surface. STM images acquired in filled-states at -2 to -3 V at I$=100$ pA.}%
\label{fig:fig6} 
\end{figure}

Higher resolution images of nascent pits 2, 6, and $>$20 atomic layers deep in Fig.~\ref{fig:fig6} (a-c) show that pits evolve following a well-known route of step-flow pinning against a point defect.~\cite{Me86} Fig.~\ref{fig:fig6} (a-b) reveal the size of the pit nuclei, as well as the subsequent mode of pit growth. We have plotted atomic layer height contours, $\Delta z = 1.36$ \AA, to show the pit cross-section in subsurface layers. The pit lengths and widths range from 0.8-1.2$\pm$0.2 nm in their deepest resolved layer. The $2\times1$ unit cell measures $0.77\times0.38$ nm$^2$, so it is evident that the pits must start at an atomic-scale defect. A simple sketch of pit initiation and early-stage formation via pinning of advancing S$_{B}$ steps is shown in Fig.~\ref{fig:fig6}. 

The steady-state evolution of a single pit is step-pinning and bunching with an apparent tendency toward pairing to form lower-energy D$_{B}$ steps, Fig.~\ref{fig:fig6}(c), as the pit sidewalls become steeper due to the crowding of S$_{B}$ steps on their downhill S$_{A}$ neighbors.~\cite{Ch87, A90, J99,L16} D$_{B}$ steps are favored as the angle of step bunches becomes larger than $2-4^{\circ}$.~\cite{Ch87,Sw89, A90} In the films that we have grown, the angles ($<10^{\circ}$) of the pit walls are too low to be consistent with formation of $\{113\}$ facets ($25.4^{\circ}$).

In later stages of growth, two trends become apparent, illustrated by Fig.~\ref{fig:fig6} (d). First, clusters of pits form as multiple steps wrap around more than a single pit at once. Second, an intuitive smoothing effect on remaining ledges between the pits becomes apparent.~\cite{M10s} As pits become more dense, the remaining ledges between pits diminish to sizes where growth is dominated by single nucleation events on ledges. As we will explain, this suppresses new pit formation because new pit nucleation sites occur at APBs between adatom islands, which promote oxygen uptake. Steps flow to the pit edges producing relatively step-free regions between pits. Such an effect could be useful for step-flow engineering of atomically-flat structures.

\subsection{\label{sec:level2}Thermodynamic metastability of pits}

\begin{figure}[h]
 \includegraphics[width=250 pt]{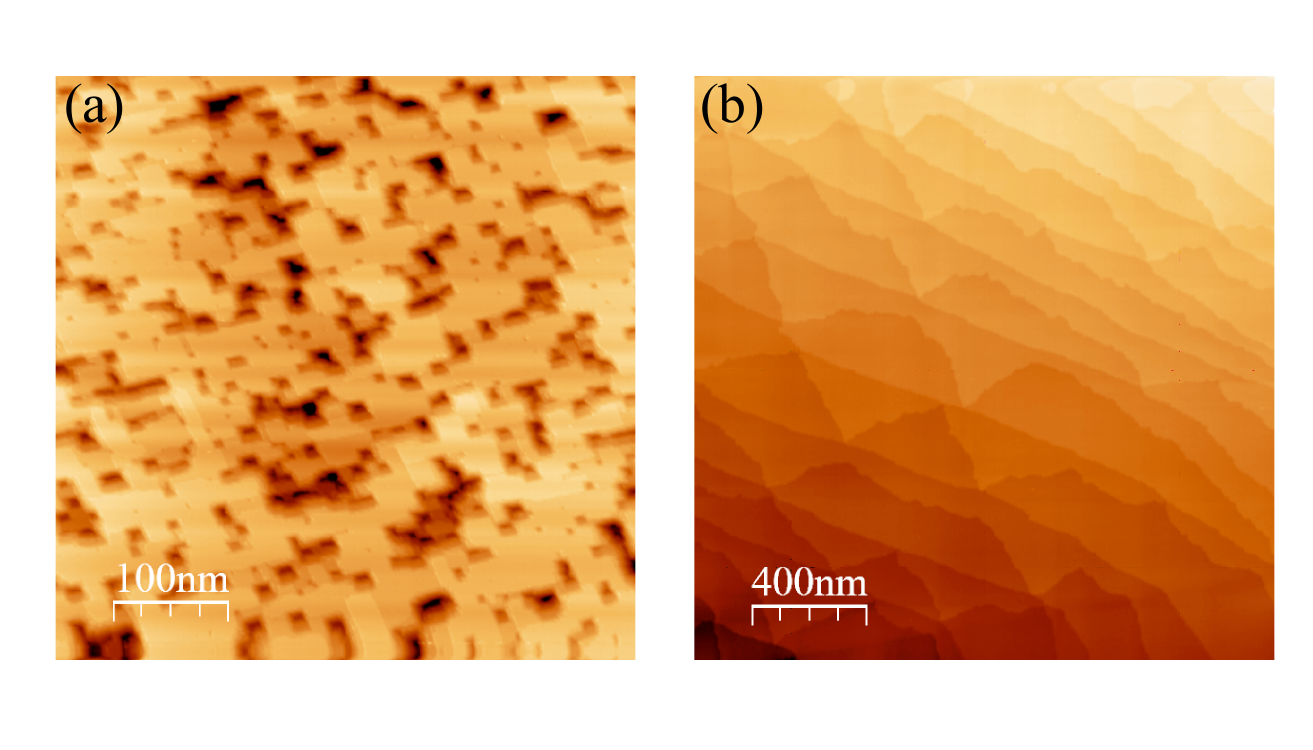}%
\caption{ Brief anneals at elevated T remove growth pits. (a) Growth pits a few nanometers deep in an 8 nm-thick film grown at $T=550^{\circ}$C.  (b) After annealing for 5 minutes at $T=750^{\circ}$C, the pits have healed and returned a surface with evenly spaced single-layer steps. These images were acquired in filled-states at -2 to -3 V at I$=100$ pA . }%
\label{fig:fig7} 
\end{figure}

We find that our pits are a metastable feature. Annealing a pitted sample smoothes the surface. Fig.~\ref{fig:fig7} (a) shows a 8 nm-thick $T=550^{\circ}$C growth surface, dense with few-nanometer deep pits, that after annealing for 5 minutes at  $T=750^{\circ}$C becomes smooth and uniformly stepped, Fig.~\ref{fig:fig7} (b). In relaxed homoepitaxial Si, pit formation costs the additional free energy of each step edge bounding each pit, building-in a driving force for surface diffusion mediated smoothing to fill-in the pits in an anneal that desorbs or dissolves the oxide nuclei. UHV annealing an oxidized Si(100) surface at $T\gtrsim700^{\circ}$C for a few minutes will remove nanometer-thick oxide layers.~\cite{Sh86}

By contrast, growth pits form as a {\it  thermodynamically spontaneous} intrinsic feature of heteroepitaxial growth of both compressive and tensile strained (Si/SiGe or SiGe/Si) thin films under deposition conditions similar to ours.~\cite{Pi92,C95, Di00,Gray02} For example, compressively strained Si$_{70}$Ge$_{30}$ deposition on Si (0.9 \AA/s, T=$550^{\circ}$C) produces distinct pits that are evident by film thicknesses $\sim$15 nm.~\cite{Gray02} 3D pits provide energetically equivalent strain-energy relief to more widely studied 3D islands (quantum dots).~\cite{Te94} And in contrast to our annealing results, strain-driven pits actually enlarge during annealing.~\cite{V98,J07}

\section{\label{sec:level1} Discussion}

\subsection{\label{sec:level2} Role of oxygen and growth antiphase boundaries in pit nucleation}

We have demonstrated that growth pits in Si epitaxial thin films are vastly more probable (1) with increasing oxygen-bearing contamination in the growth environment (Figs.~\ref{fig:fig2}-\ref{fig:fig4}), and (2) when 2D adatom islands are present during growth (Fig.~\ref{fig:fig5}). From these observations, we conclude that pits nucleate at oxide defects that occur at surface features unique to the 2D islanding growth mode. 

Because the ordered Si(100)-$2\times1$ surface is semiconducting, defects such as atomic steps, vacancies, or impurities that induce states closer to $E_f$ are the most probable reaction sites for chemisorption processes.~\cite{Av91,Br99} On Si(100)-$2\times1$ surfaces with clean step-terrace structure, the oxygen sticking coefficient is $<<1$ and oxidation initiates at point defects~\cite{Av91} and step-edges.~\cite{Br99}  Beyond obvious additional length of atomic step edge bounding each island, a surface defect distinct to island growth mode is the intrinsic APBs that occur with $50\%$ probability between adjacent Si islands. 

An APB occurs at the junction of two surface domains, e.g. islands, whose dimer reconstructions are out-of-phase by $1/2$ of the $2\times1$ unit cell perpendicular to their dimer rows.~\cite{Ha89} Two islands may converge at either S$_{A}$ or S$_{B}$ edges forming A or B-type APBs.  B-type APBs rapidly capture Si adatoms and serve as preferential nucleation site for an overlayer island, and they are most often observed to be covered by an overlayer.  Hence, in low-T epitaxy, B-type APBs contribute to roughening, and most likely eventual amorphization by capturing adatoms and promoting nucleation. By contrast, A-type APBs do not preferentially capture adatoms and they are frequently observed on surface in the 2D island growth mode, see Fig.~\ref{fig:fig5} inset at $T=360^{\circ}$ C. In clean Si epitaxy, A-type APBs are entirely a surface defect, and they eventually fill-in during growth with no lasting consequences on the bulk Si structure.

Both in MBE and CVD (disilane), B-type APBs serve as reactive sticking, nucleation, and growth sites.~\cite{Ha89,Br93,v01}  
A structural model for B-type APBs, developed from STM images, consists of two S$_{B}$ edges and a row of split-off dimers,  providing relatively high densities of dangling bonds and sites with strained geometry unique from the the surrounding $2\times1$ reconstruction, making the APBs preferential sticking sites.~\cite{Ha89,Br93} Scanning tunneling spectroscopy measurements show that the S$_{B}$ steps have enhanced DOS at E$_F$ as compared to S$_{A}$ steps.~\cite{Av91} 

Based on our observations about the role of oxygen and 2D Si islands in pit formation, our hypothesis is that pits heterogeneously nucleate at \AA-sized oxide step-pinning centers that form preferentially on B-type APBs. Subsequently, passing steps pin against the oxide features. We have not been able to resolve the naked pit nucleus by STM during growth, but it is likely to consist of an \AA-sized oxide-complex inserted at a point that stabilizes the local reconstruction, adequately delaying the interlayer bond recoordination required to go from the $2\times1$ reconstruction to a bulk diamond structure. The preferred point for oxygen binding is direct oxygen insertion into the surface dimer bond.~\cite{En93}  Naturally, since Si-O bonding is stronger than Si-Si bonding, we speculate that oxygen insertion in a dimer-bond at an APB adequately delays bulk bonding for a period long enough to allow a step pile-up and pit formation (for our choice of growth conditions). 

Our findings are reminiscent of oxide-induced 3D mound formation by step-pinning defects during Si etching by molecular oxygen.~\cite{Do93, Wu94, Se94,Se95,Br99} Comparison of our results with step-dynamics during oxidation-induced etching processes can be useful since etching is effectively time-inversion of growth. During oxygen-induced etching of both the Si(100)-$2\times1$ and (111)-$7\times7$ surfaces, the surface roughens by step-pinning at nanosized oxide defects to produce multilayer islands~\cite{Do93, Wu94, Se94,Se95,Br99} During etching, Si flux is away from the surface and steps are retracting, while during epitaxy steps are advancing, hence step-pinning during etching produces mounds, while during epitaxy it produces pits. Oxygen also drives step-pinning, step-bunching, and faceting on other surfaces, both with and without growth flux.~\cite{Ab68} 

During Si CVD growth from silanes, it is possible that pits may form by the mechanism described here, although at lower densities since CVD growth surfaces will be relatively protected from oxygen adsorption by partial H passivation under typical growth conditions.~\cite{G93,Br93,Wu98} The likelihood for pitting will increase with growth T and diminishing growth flux, both of which diminish the H passivation, exposing more oxygen binding sites and increasing the potential for pit formation. STM images of growth surfaces during CVD via silanes are qualitatively similar to MBE surfaces, with islands and APBs, suggesting the possibility for oxide and pit nucleation.~\cite{Br93,Wu98} 

\section{\label{sec:level1}Summary}
Our findings more completely elucidate the formation mechanism of growth pits in Si(100) homoepitaxy and reveal an undetected missing ingredient in reports explaining pits as intrinsic features of Si homoepitaxy. Specifically, we show that oxides cause pit formation. We show that pit formation is another manifestation of familiar step pinning effects common in growth, sublimation, ion sputtering, and etching on many solid crystalline surfaces.   We show that pits start at \AA-sized step-pinning defects, most likely molecule-sized oxide complexes, that nucleate preferentially on APBs. Pits deepen by subsequent step bunching around pit nuclei.  A steady-state growth condition is reached when the surface is so dense with pits that the remaining terraces support only single nucleation events, thereby eliminating new APBs and slowing oxide defect formation. Oxygen-induced pits are metastable and vanish into the original vicinal stepped surface during anneals hot enough to remove oxide defects.  Finally, these results reemphasize the importance of carefully controlling oxygen and oxides in the growth environment, and deposition sources, owing to the facile robust oxidation of Si surfaces.

\begin{acknowledgments}

We would like to acknowledge useful discussions with John Reno (Sandia) and Chris Richardson (Laboratory for Physical Sciences, University of Maryland). This work was performed, in part, at the Center for Integrated Nanotechnologies, an Office of Science User Facility operated for the U.S. Department of Energy (DOE) Office of Science. Sandia National Laboratories is a multimission laboratory managed and operated by National Technology and Engineering Solutions of Sandia, LLC., a wholly owned subsidiary of Honeywell International, Inc., for the U.S. Department of Energy's National Nuclear Security Administration under contract DE-NA-0003525.
\end{acknowledgments}

\bibliography{ref.bib}

\begin{thebibliography}{60}
\bibitem [{EBB()}]{EBB17}%
  \BibitemOpen
  \href@noop {} {}\bibinfo {note} {We use the notation SA and SB to denote
  single monolayer steps with dimer bonds perpendicular and parallel to the
  step edge, and DB to denote double-height rebonded B-type steps.}\BibitemShut
  {Stop}%
\end{thebibliography}

\end{document}